# Influence of carbon nanotubes parameters, boundary conditions, and scaling of the represetative volume on percolation and electric permeability of a nanocomposite

N.A. Gudkov[1], S.V. Lomov, I.S. Akhatov, S.G. Abaimov

*CREI Centre for Design Manufacturing and Materials, Skolkovo Institute of Science and Technology, Bolshoy blvd., 30, bld.1 Skolkovo, 121205 Moscow, Russian Federation*

## Abstract

The paper aims at the development of a digital twin modelling the electrical conductivity of multi-wall carbon nanotube (MWCNT)/polymer nanocomposite. A comparative analysis is conducted for various formulations of boundary conditions in the representative volume element (RVE) with uniform and isotropic distribution of CNTs. Three types of boundary conditions studied: uniform boundary potentials for an RVE, uniform boundary potentials for a chain of two RVEs, and periodic boundary potentials. It is demonstrated that to calculate the electrical conductivity of the material correctly, the implementation of periodic boundary conditions for the RVE is required; on the contrary, the use of uniform boundary potentials leads to significant overestimation in conductance. The sensitivity of average conductivity to the RVE size is determined for different volume fractions (Vf) of CNTs. For example, the 5 μm size of the RVE is found to be optimal for 1% Vf of CNTs with 50 nm diameter and 5 μm average length. Beside the RVE size, we verify how model output changes in response to tuning parameters of the CNT geometry, such as curvature and torsion (twist). We find the presence of the second parameter, torsion, to be imperative for correct conductivity modelling since usage of the curvature alone leads to unavoidable dependence on segment size during CNT discretization. In literature, the influence of intrinsic MWCNT resistivity is often discarded as negligible due to ballistic transport of electrons. To verify this assumption, we check the sensitivity of results to finite resistivity of CNTs. Percolation threshold and the critical index are calculated for different sets of model parameters. The model finite-size effect on the percolation threshold is investigated.

## 1. Introduction

Conductive polymers possess broad applicability in industry [1,2]. CNTs, both single-wall (SWCNTs) and multi-wall (MWCNTs), present a promising opportunity to provide electrical conductivity to polymers because of their high intrinsic electrical conductivity [3], 1D geometry, and high aspect ratio leading to low percolation threshold [4]. To design a material with required functional properties, the development of a digital twin is often necessary to avoid extensive experimental work. The digital twin must take into account the physical aspects of the material to be modelled and be based on fundamental physical grounds.

The first phenomenon that imparts electrical conductivity to nanocomposite is intrinsic conductivity of CNTs provided by the ballistic transport of electrons along them. CNTs possess semi-conductive or metallic properties depending on their diameter, chirality, and number of walls [5]. It is near to impossible to control the chirality parameter at manufacturing, but it determines conductivity only of SWCNTs and becomes irrelevant in case of MWCNT which are the target of our research.

---

[1] Corresponding author at: *CREI Centre for Design Manufacturing and Materials, Skolkovo Institute of Science and Technology, Bolshoy blvd., 30, bld.1 Skolkovo, 121205 Moscow, Russian Federation*
*e-mail: nikita.gudkov@skoltech.ru*

As shown by experiments, specific electrical resistivity of MWCNTs is expected to be in the range of $7 \times 10^{-5}$ to $4 \times 10^{-3}$ $\Omega \cdot cm$ [6]. The polymer matrix plays the role of an insulator with high volume resistivity of the order of $10^{15}$ $\Omega \cdot cm$ [7]. CNT/polymer nanocomposites can be produced with specific resistivity as low as $3 \times 10^{1}$ $\Omega \cdot cm$ [8] but for practical values of additive volume fractions resistivity of $10^{7} \div 10^{5}$ $\Omega \cdot cm$ can be achieved. The electrical conductivity of nanocomposites is determined not only by the intrinsic conductivity of CNTs but mostly by the transport of electrons between them. When two CNTs are not in direct contact with each other, the transport is provided by the tunnelling of electrons over small distances. The dependence of tunnelling resistance on the distance can be calculated using the Landauer-Büttiker formula [9], implemented in the present research. Neglecting tunnelling phenomenon, as a possibility to form an electrical chain at finite distance between CNTs, leads to a significant underestimation of electrical conductivity for a nanocomposite [10]. In practice, since the resistivity of ballistic transport is much lower than the tunneling resistivity at joints, the intrinsic resistivity of MWCNTs is often counted as zero. In our study, we verify this assumption.

Another important aspect in the modelling is how geometry of CNTs influences percolation. In the theory of percolation, analytical approaches are generally available for discrete lattices [11,12], not for continuous systems like assemblies of CNTs of variable curvature. In the latter case, numerical simulations are required [10,13–19]. In early studies, models contained straight CNTs [13,20–25]. In more recent publications, the curvature of CNTs was taken into account leading to the significant increase in the percolation threshold [10,17]. The percolation threshold has been shown to depend on the geometrical features of the CNT assembly which include but are not limited to orientation distribution, aspect ratio, agglomerates size [14,18]. The critical conditions for conductivity have been studied, for example, in [26–28]. These critical conditions and the conductivity dependence on the CNT volume fraction are important if the nanocomposite is used as a deformation sensor, as the sensitivity of such a sensor is maximal near the percolation threshold [29,30].

Averaged electrical conductivity of a nanocomposite which is heterogeneous is determined by homogenization. Representative volume element (RVE) is used for homogenization in many works on modelling of CNT nanocomposites [10,13–19]. The essence of this approach is that infinite medium behavior is represented by the RVE of finite size. The arrangement of CNTs within the RVE is statistical. The matrix is often considered as non-conductive and nodal analysis of conductivity of CNT architecture is performed. In [31,32], the analytical model of an RVE conductivity is developed and applied to hybrid composites, containing both fibres and CNTs.

In the majority of studies [13,14,16–19], the average conductivity of the RVE is calculated by applying a uniform boundary potential on one RVE face and another boundary potential, which is uniform as well, on the opposite face of the RVE. This is, in fact, equivalent to the case where one attaches absolutely conducting plates to the mentioned faces and applies voltage not to the RVE directly but to the plates. This technique returns correct conductivity values only when the RVE size is large. However, the purpose of an RVE introduction is to substitute numerical simulations of an infinite medium with the model of finite size, and the smaller the size of an RVE, the better the computational efficiency. However, in case of limited RVE size, the implementation of uniform potentials on RVE faces may lead to the overestimation of the electrical conductivity of the nanocomposite [10,15]. The source of error can be illustrated by the fact that the conductivity of a nanocomposite is represented by an infinite medium while an RVE presents only a small part of this medium. Imagine an infinite medium segmented into a grid of RVEs so that the problem transforms into the task to find conductance of the assembly of RVEs. Introduction of the absolutely conductive plates at RVE faces creates many non-existing connections in the nodal analysis of this medium leading to the overestimation of the resulting conductivity. It is as if we have manufactured many cubic samples and putting them together to form an infinite medium, but before doing that covered opposite faces of cubes with silver paste. In so formed medium, silver paste generates multiple undesired connections, significantly boosting the conductivity in comparison to a real medium.

The present paper explores alternative boundary conditions and demonstrates that the true periodic boundary conditions for the RVE are necessary, advancing the previous publications [33,34]. Beside the influence of model parameters, we investigate behavior of the conductivity simulation and the percolation threshold in particular, with scaling of the RVE.

The paper is structured as follows: Section 2 describes the procedure of geometry generation and nodal analysis of the connectivity graph. Section 3 presents the sensitivity of the model to the size of RVE, curvature, twist, and aspect ratio of CNTs. Section four studies scaling phenomenon in percolation, in particular, a critical index of electrical conductivity, and investigates the finite-size effect on the percolation threshold.

## 2. Simulation procedure

The simulation procedure for calculating electrical conductivity consists of the following steps:

- – CNT geometry generation;
- – Analysis of the CNT network connectivity;
- – Calculation of the electric conductivity (nodal analysis).

### 2.1. CNT geometry generation

The algorithm generates the geometry of curved CNTs with spatially uniform and orientation-isotropic distribution. A CNT grows in discrete steps from randomly placed seeds in a random direction until its targeted length is reached. This approach is based on the algorithm from [10,17,18,35–38], with a modification that we control not only curvature but also torsion of a 1D CNT centerline in the 3D space.

The input parameters of the geometry generator are the RVE size $a_{RVE}$, CNT geometry characteristics such as its diameter $D_{CNT}$ and length distribution, the CNTs volume fraction $Vf$, and the parameters, which define the waviness of CNTs: maximal angle of deflection $\alpha_{curv}$, maximal angle of torsion $\beta_{tors}$, and the growth segment length $l_{seg}$. The actual angle of the deflection $\alpha$ at every CNT growth step is random, chosen from the range $[0, \alpha_{curv}]$, as well as twist angle $\beta$ from the range $[-\frac{\beta_{tors}}{2}, \frac{\beta_{tors}}{2}]$ (Figure 1a). Building successive CNT segments according to the algorithm parameters is termed "growth" below.

The maximal curvature and maximal torsion are related to the maximal angles of deflection and torsion via the following equations:

$$\rho = \frac{2 \sin\left(\frac{\alpha_{curv}}{2}\right)}{l_{seg}} \tag{3}$$

$$\tau = \frac{\beta_{tors}}{2\pi l_{seg}} \tag{4}$$

Note that if $\beta_{tors} = 2\pi$, then the torsion is not limited.

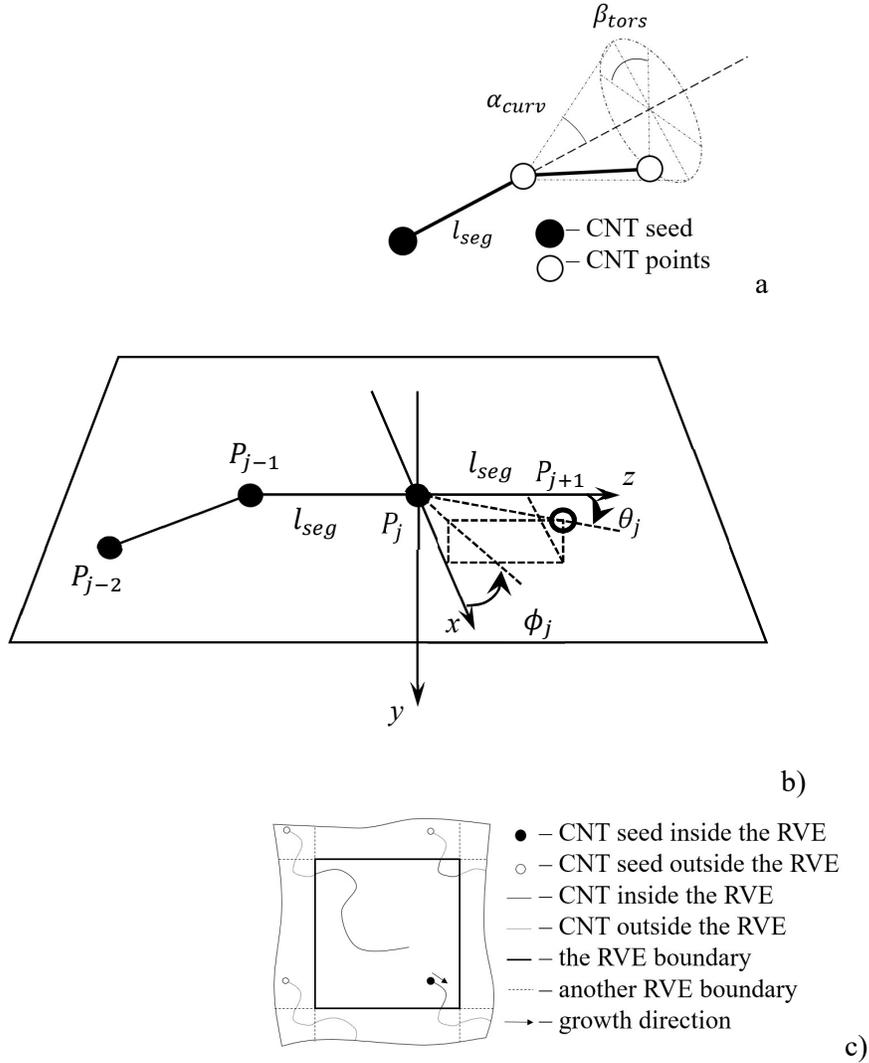

Figure 1 CNT geometry generation: (a) angles limiting a CNT section curvature and torsion; (b) growth of a new $j$-th segment; (c) growth of a CNT across the RVE boundaries

Once a CNT seed is generated, the direction of growth of the first CNT segment is chosen randomly and is described by spherical angles $\phi_1$ and $\theta_1$ in the global coordinate system. For the orientation distribution to be uniform, the angles are chosen according to

$$\phi_1 = \pi(2\xi x_1 - 1)$$
$$\theta_1 = \arccos(2\zeta_1 - 1)$$

(1)

where $\phi_1$ and $\theta_1$ are angle coordinates, $\xi$ and $\zeta_1$ are non-correlated random numbers uniformly distributed in $[0,1]$.

The CNT grows in the chosen direction by the length increment $l_{seg}$. With the points $P_{j-2}$, $P_{j-1}$, and $P_j$ constructed, the next point $P_{j+1}$ is chosen as illustrated in Figure 1b. A local coordinate system $xyz$ is defined as follows: axis $z$ corresponds to the direction $P_{j-1}P_j$; axis $x$ is normal to $z$ and lies in the plane defined by three points $P_{j-2}, P_{j-1}$, and $P_j$, directed towards $P_{j-2}$; axis $y$ is normal to $x$ and $z$. The direction of the segment $P_jP_{j+1}$ is defined by the spherical angles $\phi_j$ and $\theta_j$ in the local coordinate system $xyz$. Angle $\phi_j$ is measured from axis $x$ in plane $xy$ and is uniformly distributed in range $[-\frac{\beta_{tors}}{2}, \frac{\beta_{tors}}{2}]$. Since $x$ is directed towards $P_{j-2}$, measuring angle $\phi_j$ from axis $x$ allows us to slowly rotate the plane where CNT grows with the given curvature. For example, taking $\beta_{tors} = 0$, we generate a 2D CNT, growing in a particular plane by coils. For non-zero $\beta_{tors}$, this plane will be continuously changing, leading to the emergence of a 3D pattern. Angle $\theta_j$ is measured from axis $z$ and is uniformly distributed in range $[0, \alpha_{curv}]$, providing slowly changing curvature. In the result, the new segment $P_j P_{j+1}$ lies within the sector of angle width $\beta_{tors}$ of the cone of angle $\alpha_{curv}$ whose axis coincides with the previous segment $P_{j-1} P_j$. The values of angles are chosen by:

$$\phi_j = \frac{\beta_{tors}}{2} * \left(2\xi_j - 1\right)$$
$$\theta_j = \arccos\left(cos(\alpha_{curv}) + \zeta_j(1 - cos(\alpha_{curv}))\right) \qquad (2)$$

where $\xi$ and $\zeta_j$ are non-correlated random numbers uniformly distributed in [0,1]. The next segment of the CNT has the same length $l_{seg}$.

The generation of the CNT points continues until the full length of the CNT is reached. CNTs are generated until the given volume fraction of the CNT assembly is reached.

An RVE is treated as an element of periodicity of the infinite medium. Hence the medium is represented as an infinite array of the identical volume elements, created by translation of the RVE. Every seed that is generated within the RVE represents seeds in every cell of the array. When a CNT grows during its creation by the algorithm, it can cross the RVE boundary to be continued into the adjacent cell to keep the periodicity of the geometry. This means that a CNT leaving the RVE at one face immediately returns to the RVE at the opposite face. If CNT length is much larger than the RVE size, CNT may cross RVE's boundaries many times, every time coming back from the opposite boundary at the same location. Figure 1c illustrates this process, for simplicity in two dimensions.

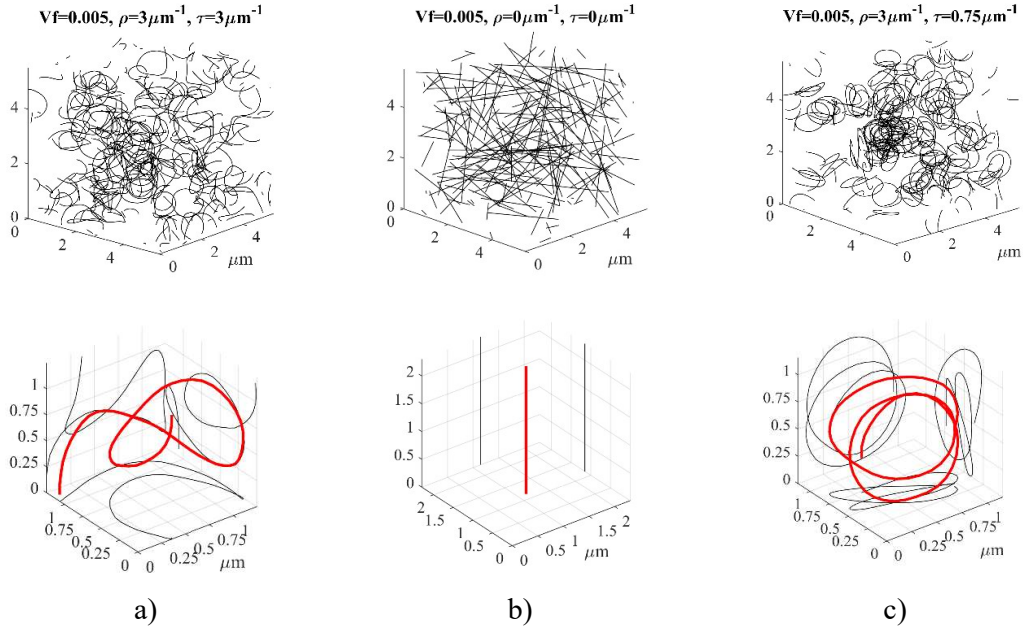

a)                              b)                              c)

Figure 2. RVE (top) and one CNT sample (bottom) with different geometry parameters: a) default geometry (Table 1. Input parameters for simulation.), b) $\rho = 0\ \mu m^{-1}$, $\tau = 0\ \mu m^{-1}$, c) $\rho = 2\ \mu m^{-1}$, $\tau = 0.75\ \mu m^{-1}$. CNT samples are shown in insets.

The full length is chosen randomly for every CNT in accordance with the Weibull probability distribution function [39,40] with parameters $a$ ("shape parameter") and $b$ ("scale parameter"):

$$f(x; a, b) = \begin{cases} \dfrac{a}{b}\left(\dfrac{x}{b}\right)^{a-1} e^{-\left(\frac{x}{b}\right)^{a}}, & x \geq 0 \\ 0, & x < 0 \end{cases} \tag{5}$$

The CNT geometry is interpolated by straight lines between the segment points. A list of CNTs is the result of the geometry generation algorithm. Items in the list are $\left((x_j, y_j, z_j)\right)^{i}$, where $i$ is the number of a CNT to which the $j^{\text{th}}$ point $(x_j, y_j, z_j)$ belongs, and $j = 1 \ldots N_i$, where $N_i$ is the number of points in $i^{\text{th}}$ CNT. The algorithm allows generalization to generation of non-uniform spatial and orientation CNT distributions.

## 2.2. Analysis of the CNT network connectivity

During this step of computational procedure, points where nanotubes interconnect with each other are defined. The main connection criterion is the distance between the CNTs being smaller than the maximal distance of tunneling $d_{cutoff}$ [15]. Note that $d_{cutoff}$ is the distance between CNT surfaces (not center lines).

List of CNTs geometry points are obtained after geometry generation and utilized here as an input together with the CNT diameter and $d_{cutoff}$.

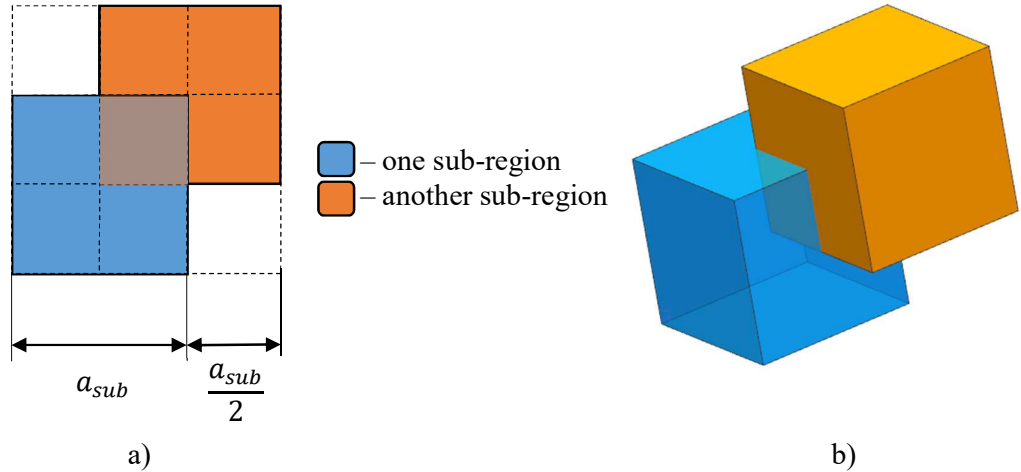

a)　　　　　　　　　　　　　　　　　　　　　b)

Figure 3. 2D (a) and 3D (b) representation of sub-regions in the RVE

To speed-up the search for the neighboring points on the CNTs, the RVE is divided into mutually intersecting sub-regions in a such way, that each point of the RVE belongs to eight intersecting sub-regions (Figure 3). The size of the search sub-region $a_{sub}$ is a computational parameter, adjusted for minimal computational time. For each sub-region, the CNTs which have at least one point inside it are defined and algorithm searches connections only between these nanotubes; then the next sub-region is considered. A similar approach was described in [17]. As sub-regions cross each other as shown in Figure 3, no connection is missed, if $\frac{a_{sub}}{2} > d_{cutoff} + D_{CNT}$.

Since CNTs are represented as nodes (filled dots in Figure 3) and straight segments between them (solid black lines), the analysis goes over all pairs of segments. For a pair of segments, one belonging to the first CNT, another belonging to the second CNT, the shortest distance between them in 3D space is found by the coordinates of their four nodes (yellow lines). If one segment has more than one connection to another CNT, only the shortest of them is left (solid yellow line) with others discarded (dash yellow lines). For the remaining connection, if it comes to the inner point of the segment, an additional node is created (empty circles). The segments in the pair can belong to different CNTs or to the same one, in the last case, only if they are not neighboring.

An output of this algorithm is a list of connections which consists of numbers $i$ and $j$ of CNTs connecting one another, numbers $k, l$ of points on these CNTs, and a distance between CNTs in the connection $d$ (tunneling distance) – see Figure 3. Hence an item in the list is $(i, k, j, l, d)$.

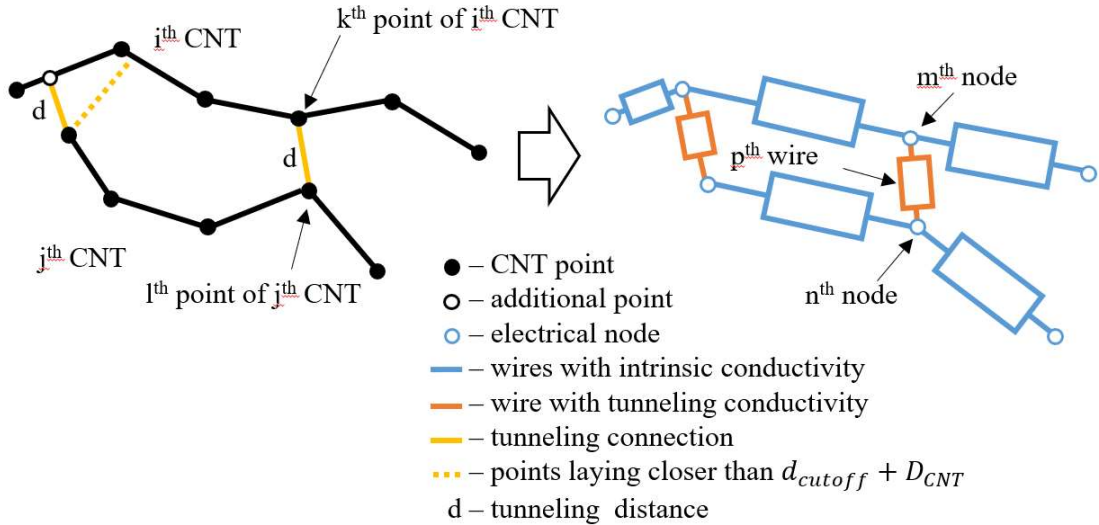

Figure 4. Connections of two CNTs

## 2.3. Calculation of electrical conductivity

In the last step of the calculation, electrical circuits are formulated, system of linear equations for nodal analysis is assembled and solved, and electrical conductivity of an RVE is obtained.

The input for this step consists of a list of connections (see the previous section) and electrical characteristics of CNTs (specific intrinsic conductivity $g_{int}$, a height of a potential barrier in connections $\Delta E$, and channel number $M$ that depends on the number of CNT walls).

The geometrical representation of CNTs is transformed into an electrical circuit representation, which consists of nodes and wires. A node is characterized by the node number $m$, its coordinates $x, y, z$, the number $i$ of the nanotube this node belongs to, and the number of point $k$ along this nanotube: $(x_m, y_m, z_m, i, k)_m$. Note that all nodes coincide with the CNT geometry points because additional geometrical points were added earlier at the connections.

The nodes on each CNT are

- the beginning and the end of the CNT;
- points at connections with other CNTs (defined by the connectivity algorithm);
- points where the CNT and a face of the RVE intersect;

A wire is described by numbers $m$ and $n$ of two nodes, connected by it, and by its electrical conductance $G$: $(m, n, G)$ (Figure 4). As it was mentioned in the introduction, CNT/polymer nanocomposites conduct electricity via intrinsic conductivity of nanotubes and electron tunneling between nanotubes, so there are two types of wires: the first simulates intrinsic conductivity of CNTs and the second simulates tunneling connection between CNTs.

To calculate intrinsic conductivity we calculate the conductivity of $p^{\text{th}}$ wire as:

$$G_p = \frac{g_{int}\pi D_{CNT}^2}{4 l_p} \tag{6}$$

where $g_{int}$ is the specific intrinsic conductivity, $D_{CNT}$ and $l_p$ are the diameter of the CNT and length of the $p^{\text{th}}$ wire. The length of a wire is calculated not as a shortest distance between two points it connects but as a distance along the CNT to which these two points belong.

To obtain tunneling conductance $G_{tunnel}$ between CNTs, Landauer-Büttiker formulae are used [18]:

$$G_{tunnel} = \frac{2e^2}{h} MT, \tag{7}$$

$$T = \begin{cases} exp\left(-\dfrac{d_{vdW}}{d_{tunnel}^{(1)}}\right) if \quad 0 \leq d \leq D_{CNT} + d_{vdW}, \\[2mm] exp\left(-\dfrac{d - d_{CNT}}{d_{tunnel}^{(1)}}\right) if \ D_{CNT} + d_{vdW} < d < D_{CNT} + d_{polymer}, \\[2mm] exp\left(-\dfrac{d - d_{CNT}}{d_{tunnel}^{(2)}}\right) if \ D_{CNT} + d_{polymer} < d < D_{CNT} + d_{cutoff}, \end{cases} \tag{8}$$

$$d_{tunnel}^{(1)} = \frac{\hbar}{\sqrt{8 m_e \Delta E_1}}; \quad d_{tunnel}^{(2)} = \frac{\hbar}{\sqrt{8 m_e \Delta E_2}} \tag{9}$$

where $T$ is the transmission probability, $d$ is the distance between nanotubes in a connection point, Planck constant $h = 2\pi\hbar$, $e$ is the electron charge, $M$ is the conduction channels number, $d_{vdW}$ is van der Waals distance, $d_{cutoff}$ is the tunnelling cutoff distance, $m_e$ is the mass of an electron, $\Delta E$ is the height of a potential barrier. The main idea in this formulation is that presence of polymer between CNTs lowers tunnel potential barrier. $d_{polymer}$ is a minimal distance between CNTs that allows polymer to permeate between them.

Next step is to construct conductivity matrices for nodal analysis [41]. For $p^{\text{th}}$ wire (resistor) with conductance $G_p$, the nodal equations are formulated as

$$[K_{mn}]_p \begin{Bmatrix} v_m \\ v_n \end{Bmatrix} = G_p \begin{bmatrix} 1 & -1 \\ -1 & 1 \end{bmatrix} \begin{Bmatrix} v_m \\ v_n \end{Bmatrix} = \begin{Bmatrix} I_m \\ I_n \end{Bmatrix}_p, \tag{10}$$

where $m$ and $n$ are nodes of the resistor, $v_m$ and $v_n$ are node potentials, $I_m$ and $I_n$ are electric currents through nodes $m$ and $n$, and $K_{mn}$ is a conductivity matrix of a resistor. In accordance with Kirchhoff's law:

$$\begin{aligned} K &= \sum_p [K_{mn}]_p, \\ Kv &= I_{ext}, \end{aligned} \tag{11}$$

where $K$ is the conductivity matrix of the CNT network, $v$ is a vector of unknown nodal potentials, and $I_{ext}$ is a vector of external currents.

To calculate the average conductivity of the network, one has to apply voltage to it, calculate the total current and then derive the conductivity. The modified nodal analysis was applied to this problem to replace nodal potentials of wires with no resistivity in the vector of unknowns by unknown currents in these wires [41]. Three types of electrical schemes, realizing such a calculation, were compared (Figure 4).

Type A is the traditional scheme where voltage is applied to opposite faces of the RVE so that potentials of all electrical nodes lying on each of these faces are identical everywhere on the face. This scheme is utilized in a majority of publications [13,14,16–19] and is, in fact, equivalent to the scheme where absolutely conductive plates are attached to the RVE faces, and the voltage is applied not to the RVE directly but to these plates (blue wires on Figure 4). The nodes lying on other faces (where voltage is not applied) are connected in pairs to provide periodic boundary conditions (red wire on Figure 4). The average conductivity is calculated according to (10) in Figure 4a, where $I$ is current caused by voltage $E$ and $a_{RVE}$ is a size of RVE cube's side.

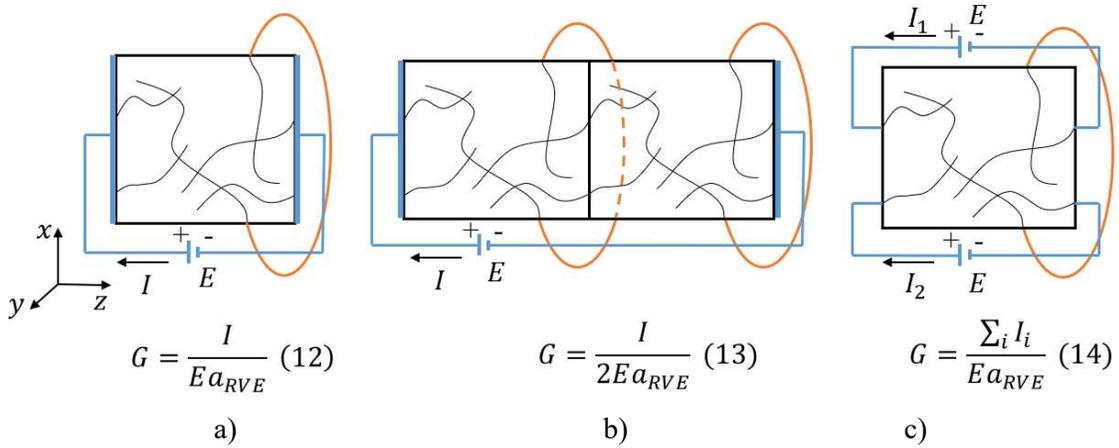

$$G = \frac{I}{E a_{RVE}} \quad (12)$$

$$G = \frac{I}{2E a_{RVE}} \quad (13)$$

$$G = \frac{\sum_i I_i}{E a_{RVE}} \quad (14)$$

a)                              b)                              c)

Figure 5. Schemes with different applications of a voltage. $G$ is the homogenized conductivity of the material: a) scheme A, b) scheme B, c) scheme C

Since scheme A relies on the assumption of uniform potentials along left and right (as in Figure 4a) RVE faces, we test its validity by the conductance of scheme B (Figure 4b), which is the assembly of two RVEs. The second RVE is identical to the first one, and due to the periodicity of the RVE structure, CNTs are passing through the boundary between two RVEs. The average conductivity is calculated according to (13) in Figure 4b. Again, uniform potentials are prescribed over the left and right faces of the assembly. If across the boundary between the RVEs we prescribed uniform potentials again, the conductivity would be exactly as in scheme A (the conductance would be twice lower). But we do not prescribe potentials on the boundary between two RVEs letting them differ one point from another and being non-uniform across this boundary. If the idea behind scheme A, that uniformity of potentials across the boundary does not influence the result, is correct, scheme B will provide conductivity identical to scheme A (twice lower conductance). On the contrary, if non-uniformity of potentials at the boundary (in this case, in between two RVEs) has significant effect on the result, predictions of scheme B will differ significantly from scheme A. For the top and bottom faces of the assembly, periodical boundary conditions are realized by connecting the corresponding nodes.

In scheme C we utilize true all-side periodical boundary conditions for the RVE, and the voltage is again applied to two opposite faces, but this time, the scheme allows the potentials to variate from one node to another along the face. Let us consider node pairs on opposite faces with coordinates $(x, y, 0)$ and $(x, y, a_{RVE})$, for conductivity calculation on z-direction (see Figure 4 for the coordinate system orientation). We apply the same difference $\Delta v$ in potentials (equal to voltage) to every pair, $v(x, y, 0) = v(x, y, a_{RVE}) + \Delta v$ but allow potential $v(x, y, 0)$ and potential $v(x, y, a_{RVE})$ to variate along the left and right faces of the RVE in Figure 4c, respectively. For lateral faces, the periodic boundary conditions are implemented in a similar way, only the difference in potentials is not applied: $v(x, 0, z) = v(x, a_{RVE}, z)$ and $v(0, y, z) = v(a_{RVE}, y, z)$. After solving linear equations, nodal potentials and unknown currents $I_1, I_2, ...$ are obtained. The average conductivity is calculated according to (12) in Figure 4c, where $\sum_i I_i$ as sums of currents through faces $(x, y, 0)$ and $(x, y, a_{RVE})$.

### 2.4. Calculation of the percolation threshold and the critical index of percolating CNT

Non-zero electrical conductivity of the system appears when CNTs form an interconnected network across the RVE at the volume fraction $Vf$ of CNTs exceeding the so called percolation threshold. The percolation threshold depends on system's size [11,12]. To illustrate why, let us imagine an infinite system with percolation threshold $Vf_C^\infty$. Cutting out of this system a volume of linear size $L$ by virtual boundaries, we create a system of the finite size with percolation threshold $Vf_C(L)$. Statistical analysis tells us that always $Vf_C(L) < Vf_C^\infty$ because it is easier to form a conducting network of finite size than its infinite counterpart.

Just above the percolation threshold, the system follows the rules of continuous phase transitions with the percolation threshold being the critical point (index C in the notations above stays for "critical"). In the result, the appearing conductivity is expected to obey the power-law dependence in an infinite system

$$K = a(Vf - Vf_C^\infty)^\beta \tag{15}$$

with $\beta$ being called the critical index. Critical indices are important because their values tell us to what universality class a system belongs in the theory of phase transitions. For percolation problems, values of critical indices can be found in [12]. For nanocomposites, experimental studies are reported in [26,42].

For a system of finite size, we expect deviations [11] from power-law (15) in the very close proximity of the percolation threshold, which can be expressed as *scaling* of the system. Analysis of the scaling allows determination of the percolation threshold for an infinite system, as illustrated in Figure 6.

Let us denote by $\Pi$ the probability that at this volume fraction $Vf$ of CNTs the percolation network is formed. For an infinite system, the probability $\Pi^\infty$ to percolate is a step function, i.e. exactly zero below the percolation threshold $Vf_C^\infty$ and is exactly unity above the threshold [11,12]. For a finite system, the finite-size effect takes place, where the step function is substituted by a smooth curve like in Figure 6 provided by scaling:

$$\Pi = \Xi\left(\frac{L}{\xi^\infty}\right), \Pi \xrightarrow[L \gg \xi^\infty]{} 0, \Pi \xrightarrow[L \ll \xi^\infty]{} 1, \tag{16}$$

where $L$ is the linear size of a system and $\xi^\infty$ is the correlation length in an infinite system, which diverges as $\xi^\infty \propto |Vf - Vf_C^\infty|^{-\nu}$ near the percolation threshold. Let us substitute this divergence into (16) and consider it exactly at the point of percolation threshold of an infinite system, $Vf = Vf_C^\infty$ (at this point $\xi^\infty \to \infty$)

$$\Pi = \Xi(0), \text{ when } Vf = Vf_C^\infty \ \forall L. \tag{17}$$

So, at $Vf_C^\infty$, $\Pi$ does not depend on system's size taking some finite value $\Xi(0)$ between 0 and 1. This provides an opportunity to find the percolation threshold $Vf_C^\infty$ of an infinite system as a point at which curves $\Pi(f, L)$ intersect all being equal to $\Xi(0)$.

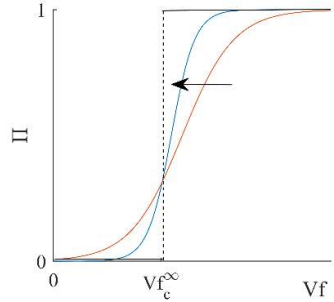

Figure 6. Finite-size effect for percolation: the colored curves correspond to the systems with finite size, arrow showing increase of this size; the black stepwise graph is for an infinite system

### 2.5. Plan of the numerical experiments and the input data

The numerical experiments were performed to investigate the following problems:

1. Influence of the modelling parameters (the model boundary conditions, the RVE size, and CNT segment size) and choice of the optimal set of the modelling parameters.
2. Influence of the CNT geometry (mean length, curvature, twist) and CNT volume fraction on the electric conductivity.
3. Influence of the non-zero resistivity of ballistic electron transport within the CNTs.
4. Influence of length, curvature, and torsion on the percolation threshold and percolation critical index

The physical parameters of CNTs and the default modelling parameters, used in these parametric studies, are shown in Table 1.

Table 1. Input parameters for simulation.

| Parameter | Symbol | Value |
|---|---|---|
| **CNT size** [15] | | |
| Weibull $a$ for CNT length | $a$ | 2.4 |
| Weibull $b$ for CNT length / mean CNT length, μm | $b$ / $\lambda$ | 5.64 / 5.0 |
| CNT diameter, nm | $D_{CNT}$ | 50 |
| **Default modelling parameters** | | |
| Maximal curvature, μm$^{-1}$ [15] | $\rho$ | 2 |
| Maximal torsion, μm$^{-1}$ | $\tau$ | 3 |
| Length of segment, μm | $l_{seg}$ | 0.333 |
| **Electric conductivity parameters** [18] | | |
| Tunneling cutoff distance, nm | $d_{cutoff}$ | 1.4 |
| Polymer cutoff distance, nm | $d_{polymer}$ | 0.6 |
| Van-der-Waals distance, nm | $d_{vdW}$ | 0.34 |
| Specific intrinsic conductivity, S·m [15] | $g_{int}$ | $10^4$ |
| Channels number [15] | $M$ | 450 |
| Height of potential barrier 1 (without polymer), eV | $\Delta E_1$ | 5 |
| Height of potential barrier 2 (with polymer), eV | $\Delta E_2$ | 3.5 |

## 3. Results and discussion

### 3.1. Choice of the modelling parameters

Figure 7a shows the influence of boundary conditions on calculated conductivity values for an RVE of size 5.5x5.5x5.5 μm. Each point on the graph is the result of averaging of 400 simulations with the RVE every time populated differently by CNTs; error bars correspond to one standard deviation. Results of scheme B deviate significantly from scheme A which demonstrates that the assumption of the uniform potentials does not pass a consistency check. When periodic BC are applied (scheme C), the results differ up to 150 times for low CNT volume fraction in comparison with the scheme A. When the volume fraction becomes higher, the difference becomes smaller, which can be explained by larger number of contacts on the RVE faces, which brings the configuration closer to uniform potentials.

Based on these results, periodic BC (scheme C, Figure 5c) is consistently used in the calculations below. The error bars show $\pm \log(e) \frac{stdev}{mean}$ values of conductivity. The scatter of the results is large for low CNT volume fraction (up to 0.01) and becomes much smaller for high volume fractions. This is natural, as the system of CNT contacts and connections is more variable, if the number of CNTs is small, hence CNT volume fraction is small.

Next, the dependency of electrical conductivity on RVE size $a_{RVE}$ is considered (Figure 7b). Results are plotted for several CNT volume fractions and demonstrate that the dependency of the apparent conductivity on the RVE size is very weak. The scatter decreases roughly linearly with increase of the RVE size. The value of 5.5 μm was chosen for the further calculations.

Figure 7c shows the dependency of calculated conductivity values on the segment length, used for the CNTs construction for default values of curvature and torsion and VF = 0.01 (Table 1). After segment size goes below 0.46 μm electrical conductivity values stabilize. The value 0.333 μm of the segment length, or 15 segments per 5 μm of the average CNT length, were chosen for the further simulations.

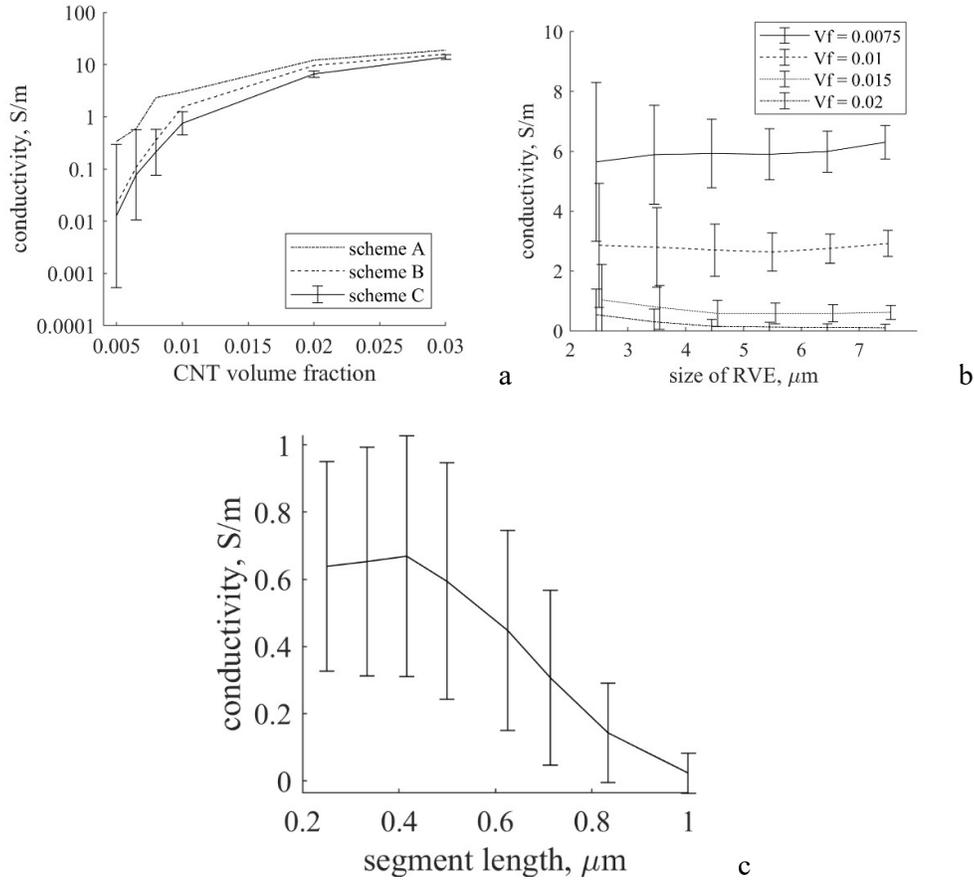

Figure 7. Influence of modelling parameters on the calculated values of conductivity: (a) conductivity vs CNT volume fraction, RVE 5.5x5.5x5.5 μm, different boundary conditions, (b) conductivity vs RVE size, different CNT volume fractions, scheme C, (c) conductivity vs number of segments used for CNTs modelling, RVE 5.5x5.5x5.5 μm, scheme C, Vf=0.01. The error bars show standard deviation in 400 simulation

### 3.2. Influence of CNT geometry and volume fraction on electric conductivity

Influence of geometry parameters, such as mean length $\lambda$ of CNTs, maximal curvature $\rho$, and maximal torsion $\tau$, on electrical conductivity is shown in Figure 8. In figs.8a-c, we plot the dependence of conductivity on volume fraction for different values of $\lambda$, $\rho$, and $\tau$. In figs.8d-f, the dependence of conductivity on $\lambda$, $\rho$, and $\tau$ is plotted for volume fraction 0.015. In figs.8g-i, the number of inter-tube connections (CNT joints) as the characteristic of the network is plotted vs $\lambda$, $\rho$, and $\tau$. Every result is an average of 400 simulations. All parameters apart from the one being investigated, are kept at their values from table 1.

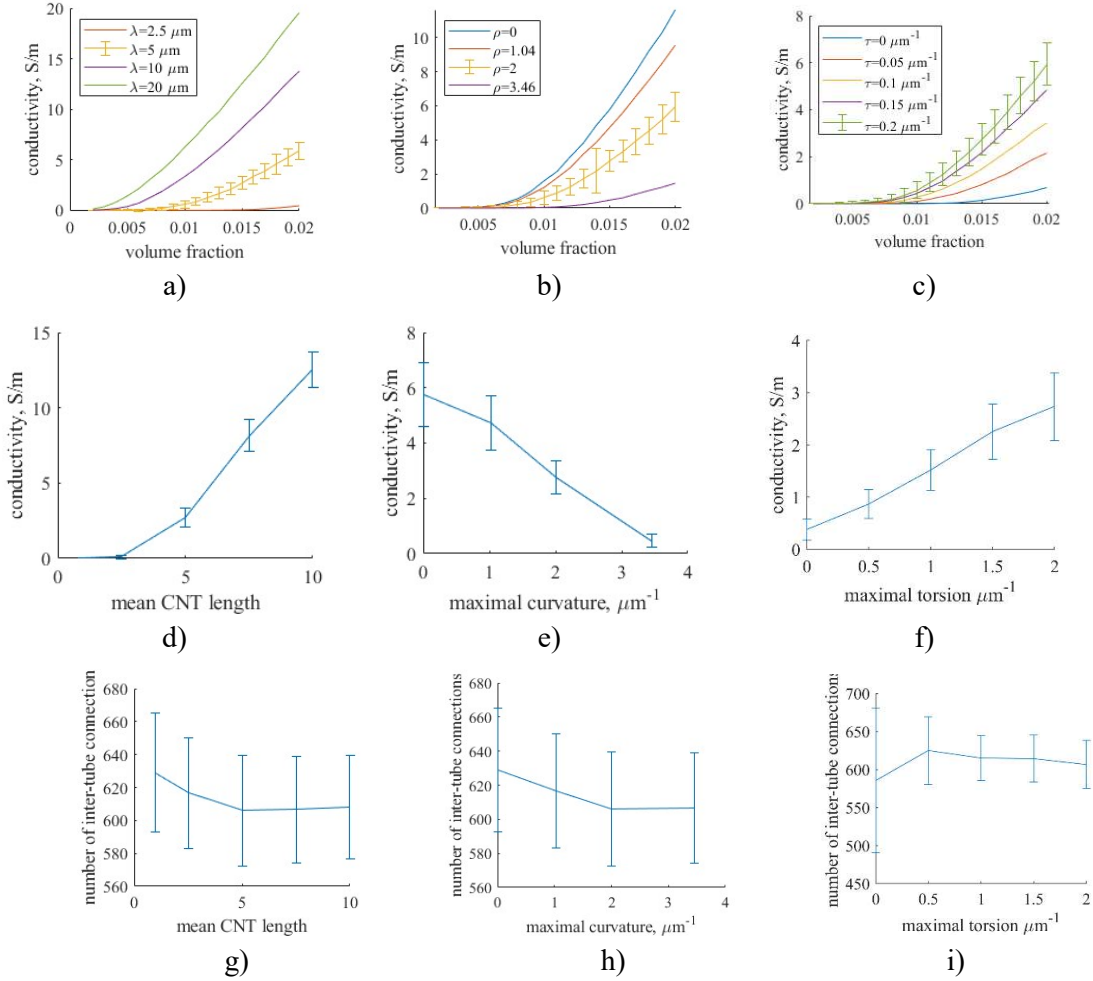

Figure 8. Influence of CNT geometric parameters on conductivity: (a) mean CNT length $\lambda$, (b) maximal CNT curvature $\rho$, (c) maximal CNT torsion $p$; (d-f): conductivity dependence on $\lambda$, $\rho$, and $p$ for volume fraction 0.015; (g-i): number of inter-tube connections as the dependence on $\lambda$, $\rho$, and $p$.

All three parameters have strong influence on CNT percolation network and, thereby, conductivity. From Figure 6g, we observe that the higher the CNT aspect ratio is, the slightly less dense the percolation network becomes (at a fixed volume fraction, when the longer the CNTs are, the lower number of them remains). However, the longer CNTs provide the longer distances connected so that, in spite of the less dense network, the conductivity increases significantly with the increase in CNT length. This trend is supported by experimental evidence [43].

From Figure 6b,e,h, we find that the increase in curvature has a strong negative effect on conductivity as expected for a transformation from straight rods to balls of threads. This result was previously observed experimentally for nanocomposite films [44]. In the mentioned publication, two films with similar filler (same diameter), but different curvature which was observed via TEM, were manufactured. Two types of samples had different antistatic effect.

The higher the torsion, the more pronounced helical structure can be formed, leading to the increase in the effective CNT diameter zone. Thereby, in Figure 6c,f we observe strong increase in conductivity in the presence of torsion, although the density of the network remains the same (Figure 6i). To the best of our knowledge, the effect of CNT torsion on conductivity has not been yet studied, and there are no experimental evidence to verify our result. We postpone its experimental verification to future work.

### 3.3. Influence of the non-zero resistivity of ballistic electron transport within the CNTs

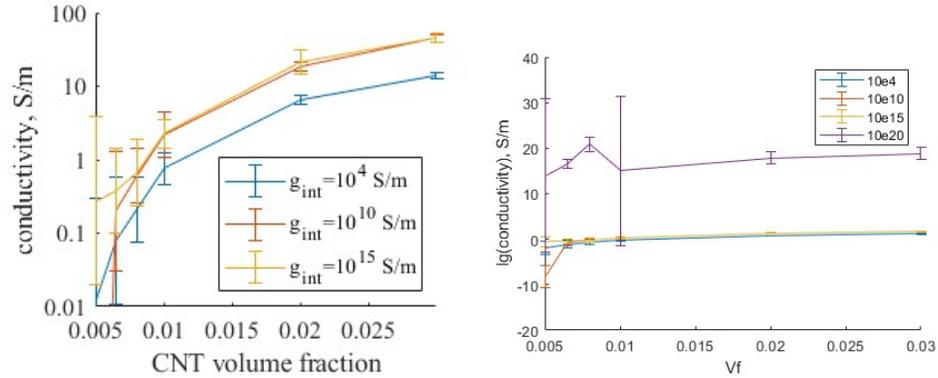

Figure 9. Conductivity vs. volume fraction for different intrinsic CNT conductivity.

In many studies, the ballistic conductivities of CNTs is taken infinite in comparison to the conductivity of the network of joints. In Figure 9, we verify this assumption by gradually increasing ballistic conductivity from $10^4$ S/m to $10^{20}$ S/m. It is demonstrated that neglecting ballistic resistivity we introduce an error around 0.5 order of magnitude.

### 3.4. Percolation threshold and the critical index

Conductivity dependence on the CNT volume fraction near the percolation threshold, eq. (15), is defined by two parameters: the percolation threshold itself and the critical index $\beta$. As it is shown in section 2.4, the percolation threshold, which depends strongly on the size of the system, should be defined using the scaling analysis; the results of such an analysis are presented in Figure 10. For four combinations of the geometrical parameters of CNTs it shows the scaling dependencies of the percolation probability $\Pi$ on the CNT volume fraction for different size of RVE. The CNT geometry cases are: the default case and three cases with one geometry parameter changed: default CNT mean length 5 μm is substituted by 10 μm in Figure 10b, default maximal curvature 2 μm⁻¹ is substituted by 1 μm⁻¹ in Figure 10c, default maximal torsion 0.2 μm⁻¹ is substituted by 0.1 μm⁻¹ in Figure 10d.

First, we observe that for all RVE sizes the probability changes in a continuous way with the CNT volume fraction, in contrast with the abrupt step-change (vertical dash line) from zero to unity in the infinite system. Therefore, in spite of the periodicity of the boundary conditions, the finite-size effect is clearly present and influences the results.

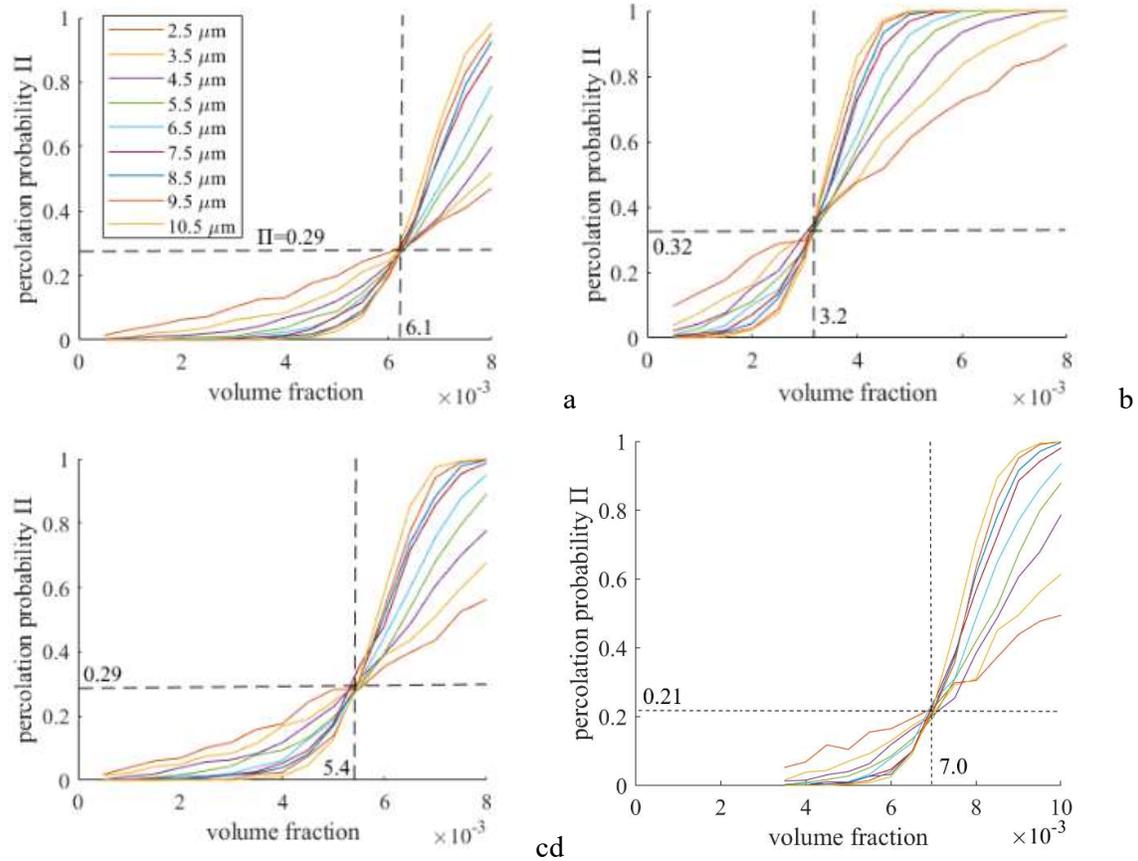

Figure 10. Probability $\Pi$ to percolate as a function of CNT volume fraction for different RVE sizes: a) default CNT geometry $\lambda = 5$ μm, $\rho = 2$ μm$^{-1}$, $\tau = 0.2$ μm$^{-1}$, b) mean CNT length $\lambda = 10$ μm, c) maximal CNT curvature $\rho = 1$ μm$^{-1}$, d) maximal CNT torsion $\tau = 0.1$ μm$^{-1}$. Vertical dotted lines show the percolation threshold of the infinite system.

Second, as it was described in section 2.5, we determine the percolation threshold of an infinite system as a point of curves intersection (vertical dash lines). For the default set of parameters it equals 0.61% CNT volume fraction. The mean CNT length has the strongest influence on it, and its doubling reduces the percolation threshold to 0.32%. CNT curvature and torsion demonstrate lower influence. Twice lower value of curvature, due to CNT straightening, reduces threshold to 0.55%. Twice lower value of torsion, due to reduction in the effective CNT diameter for joints formation, increases the threshold to 0.67%.

The differences in the percolation threshold can be explained based on the analysis of the excluded volume for the CNTs of each geometrical type. As it is known [11][45,46], increase of the excluded volume leads to decrease of the percolation threshold.

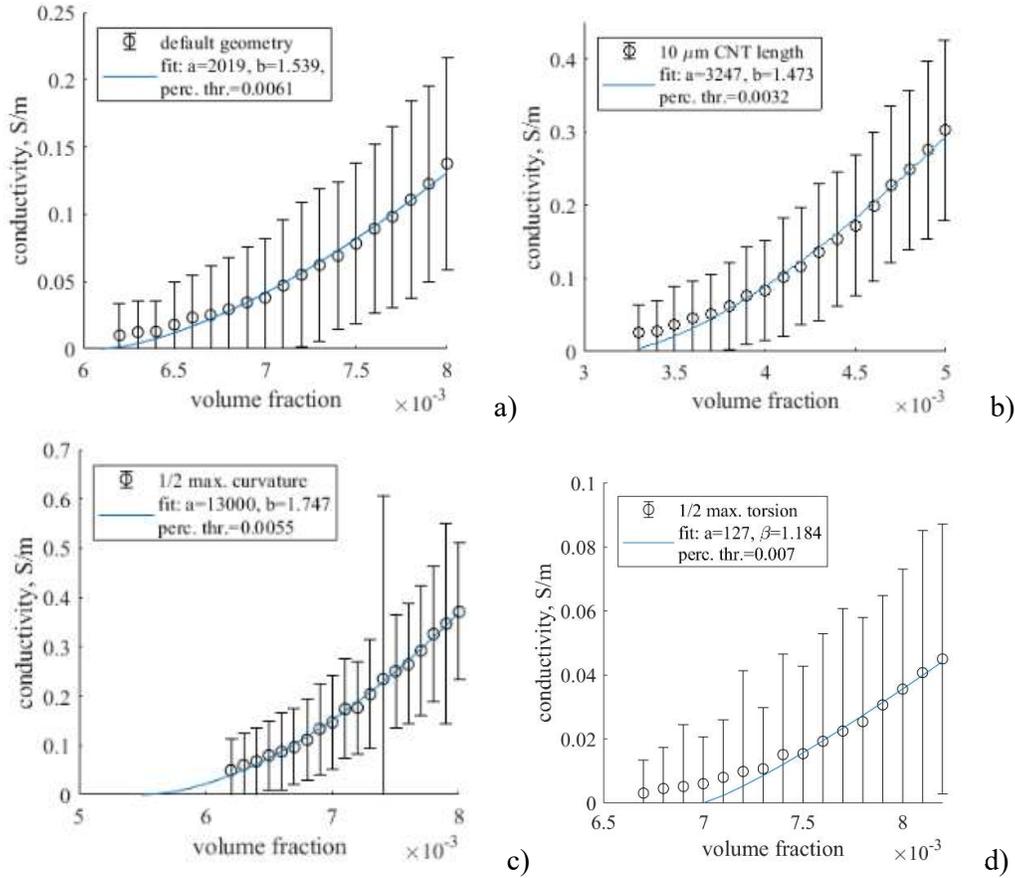

Figure 11. Conductivity vs. CNT volume fraction in the vicinity of the percolation threshold: a) default CNT geometry $\lambda = 5$ μm, $\rho = 2$μm$^{-1}$, $\tau = 0.2$ μm$^{-1}$, b) mean CNT length $\lambda = 10$ μm, c) maximal CNT curvature $\rho = 1$μm$^{-1}$, d) maximal CNT torsion $\tau = 0.1$μm$^{-1}$. The RVE size 5.5×5.5×5.5 μm$^3$.

In Figure 11, we study the influence of CNT geometry parameters on the critical index β defined by eq (15). The critical index is calculated as result of fitting of the curve mean conductivity vs volume fraction with the equation (15), the threshold values being taken from the results of the scaling analysis, described above. The used size of the RVE is 5.5×5.5×5.5 μm$^3$, which was chosen as default in the calculations in the present paper.

The critical index for all the cases lies in the range β = 1.5 – 1.7. The critical index is defined by dimensionality of the percolating network, and for 3D networks its theoretical value is 1.3 – 1.7 [47]; details of the network geometry do not matter much. Most of experimental measurements for CNT-dispersed nanocomposites show values 1.2 – 1.7 for randomly oriented CNTs [48]. There exist "outliers", with β-values of 2.7 [28] or even up to 3.9 [49]; the cause of these high values is not clear. The values calculated with the present model correspond to the theoretical range and to the most of the experimental data.

## 4. Conclusion

In our study, we have developed a digital twin for the nanocomposite with homogeneous and isotropic CNT distribution. In difference to previous works, the generator during a CNT construction controls not only its curvature but torsion also.

Our study demonstrates that application of uniform potentials at the RVE boundaries leads to a significant conductivity overestimation. Thereby, application of periodic boundary potentials is recommended.

The possibility to neglect ballistic resistivity is investigated and is demonstrated to lead to an error of half an order of magnitude.

Significant dependence of RVE conductivity and percolation threshold on CNT length, curvature, and torsion is demonstrated for a wide range of volume fractions.

The value of critical index for conductivity is found to be in the range 1.5 – 1.7 which is consistent with previous studies.

**Acknowledgements:**

This work was supported by Skoltech NGP Program (Skoltech-MIT joint project). The results have been presented at ICCS23 (23rd International Conference on Composite Structures) and MECHCOMP6 (6th International Conference on Mechanics of Composites), University of Porto, Portugal, 1-4 September 2020.